\documentclass[runningheads]{llncs}
\usepackage[T1]{fontenc}
\usepackage[utf8]{inputenc}

\usepackage{graphicx}
\usepackage{subcaption}
\usepackage{footmisc}
\usepackage{tabularx}
\usepackage{booktabs}
\usepackage{multirow}
\usepackage{array}
\usepackage{amsmath}
\usepackage{mdframed}
\usepackage{listings}
\begin{document}
% \begin{CJK}{UTF8}{gbsn} % 改为 CJK 而不是 CJK*

% Used for displaying a sample figure. If possible, figure files should
% If you use the hyperref package, please uncomment the following two lines
% to display URLs in blue roman font according to Springer's eBook style:
%\usepackage{color}
%\renewcommand\UrlFont{\color{blue}\rmfamily}
%\urlstyle{rm}
%
%
\title{Weak Supervision Techniques towards Enhanced ASR Models in Industry-level CRM Systems}
\author{Zhongsheng Wang$^{1,*}$, Sijie Wang$^{1,*}$, Jia Wang$^{2, \dagger}$,\\ Yung-I Liang$^{2}$, Yuxi Zhang$^{2}$, and Jiamou Liu$^{1, \dagger}$}
\authorrunning{Z. Wang et al.}
\institute{$^1$ The University of Auckland, Auckland, 1010, New Zealand \\ $^2$ Atom Intelligence, Hong Kong SAR, China \\
\email{\{zwan516,swan387\}@aucklanduni.ac.nz}, \email{\{johnny.wang,yoyo.liang,sharon.zhang\}@atom-intelligence.com}, \email{jiamou.liu@auckland.ac.nz}}
\maketitle              % typeset the header of the contribution

\def\thefootnote{*}\footnotetext{These authors contributed equally to this work.}\def\thefootnote{\arabic{footnote}}\

\def\thefootnote{$\dagger$}\footnotetext{Co-corresponding authors.}\def\thefootnote{\arabic{footnote}}

\begin{abstract}
In the design of customer relationship management (CRM) systems, accurately identifying customer types and offering personalized services are key to enhancing customer satisfaction and loyalty. However, this process faces the challenge of discerning customer voices and intentions, and general pre-trained automatic speech recognition (ASR) models make it difficult to effectively address industry-specific speech recognition tasks. To address this issue, we innovatively proposed a solution for fine-tuning industry-specific ASR models, which significantly improved the performance of the fine-tuned ASR models in industry applications. Experimental results show that our method substantially improves the crucial auxiliary role of the ASR model in industry CRM systems, and this approach has also been adopted in actual industrial applications.

\keywords{ASR in CRM  \and Data Augmentation \and Model Fine-tuning  \and Industrial Application.}
\end{abstract}

\section{Introduction}
A {\em Customer Relationship Management} (CRM) system is essential for managing customer interactions and data across various communication channels \cite{adlin2019current}. By centralizing customer information, CRMs improve communication, personalize service, and enhance customer satisfaction. They automate routine tasks, boosting productivity, and provide analytics that help in making data-driven decisions. An emerging trend in CRM is to incorporate voice technology, which is becoming increasingly vital for enhancing user experience and operational efficiency. For example, SuiteCRM's mobile application uses voice-to-text processing to allow users to navigate the app, search records, and record notes verbally, improving multitasking and speed \cite{mustapha2016implementing}. Similarly, the iSpeak system automates customer service interactions traditionally handled by human agents, thereby enhancing efficiency and reducing the need for direct human contact \cite{atayero2011development}. 

\begin{figure}[!htbp] % 图片环境
    \centering % 图片居中
    \includegraphics[width=0.9\textwidth]{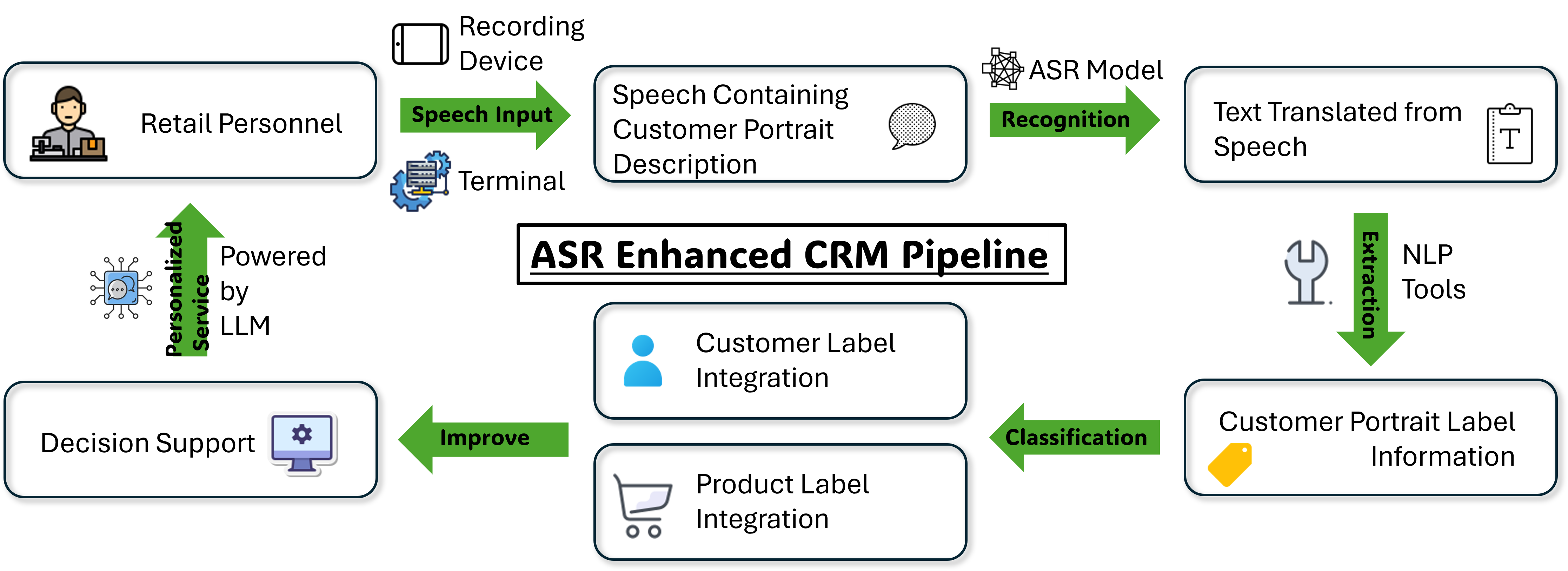} %
    \caption{A CRM Pipeline that is enhanced by voice technology and is designed for industrial deployment.} % 图片标题
    \label{fig:framework} % 图片标签，方便在文中引用
\end{figure}

Integrating voice technologies into CRM systems presents several challenges that can complicate their adoption and functionality \cite{mustapha2016implementing}. Ensuring compatibility with existing infrastructure is crucial, as CRM systems must interact seamlessly with various tools like customer databases and analytics platforms. Additionally, the technical complexity of implementing accurate {\em voice-to-text} processing is significant. These systems must handle various accents, dialects, languages, and speech patterns with high precision. Moreover, managing errors from voice recognition, which can occur due to background noise or unclear speech, without diminishing the user experience, remains a substantial challenge. Figure~\ref{fig:framework} presents our CRM pipeline where voice-to-text processing plays a pivotal role (See Section~\ref{sec:pipeline}). The design leverages {\em Automatic Speech Recognition} (ASR) to transform spoken interactions from retail personnel into actionable text data.  This pipeline not only streamlines the data collection process but also enhances the efficiency and personalization of customer service.

Although ASR models have made significant advancements in recent years, they often struggle to meet the unique needs of specific industries, hindering their effectiveness in the CRM pipeline, where precise customer portrait labels are crucial. Fine-tuning ASR models to address these industry-specific requirements is essential, but complicated by the difficulty of obtaining large volumes of accurately labeled data in real-world scenarios. The voice data recorded by sales representatives is typically unlabeled, features complex regional accents, and includes numerous proprietary brand names and colloquial terms. This low-quality data is not directly usable or easily cleaned for effective model fine-tuning, presenting a significant obstacle to improving ASR accuracy in CRM applications.

The main problem addressed in this paper is the accurate extraction of customer key features and intentions from voice inputs recorded by sales personnel, especially when general pre-trained ASR models fall short in recognizing industry-specific speech patterns. To address this problem, we designed a weak-supervision framework for fine-tuning pre-trained ASR models to meet CRM application needs. This approach leverages large language models (LLMs) and text-to-speech (TTS) models to make use of existing small, high-quality labeled datasets. With minimal human and financial costs, it generates large, high-quality datasets that can be directly used for ASR model fine-tuning. The ASR models fine-tuned with these datasets are capable of handling complex speech transcription tasks and have demonstrated superior performance. The results of experiments on various baseline models show that compared with the native model, the highest performance improvement is 63\%, and the average improvement is 51.5\%. We apply our method in real-world industry scenarios and have received positive feedback.  Our main contributions are summarized as follows:

\begin{itemize}
    \item We invented an efficient weak-supervision solution for fine-tuning ASR models in industrial CRM systems, which includes high-quality fine-tuning data generation. See Section~\ref{sec:method}.

    \item We proposed an indicator for evaluating the reasoning performance of ASR models, the \textbf{Integrated Error Rate (IER)}, which has a more objective and comprehensive evaluation significance for hybrid speech recognition tasks. See Section~\ref{sec:IER}.

    \item We deployed the ASR model fine-tuning method proposed in this paper in relevant industrial fields and verified its superiority in high-quality work performance through verification. See Section~\ref{sec:exp}.
\end{itemize}

\section{Related Work}

\noindent {\bf Voice Technologies in CRM Systems.}
The applications of voice and AI technologies in CRM systems are documented in several studies.  
%\cite{Tarokh2007} highlighted how ASR and Text-to-Speech (TTS) technologies facilitate interactive voice dialogues and automated responses. Building on this, \cite{Subramaniam2009} showed that ASR can extract business intelligence from customer voice data, improving agent productivity and insights into customer behavior.  
\cite{atayero2011development} introduces iSpeak, a voice-activated relationship management system that automates customer care services, reducing human interaction and improving efficiency. \cite{mustapha2016implementing} details the development of an Android application for SuiteCRM, incorporating speech-to-text processing for user interface navigation and database interactions, thereby enhancing usability.  \cite{adlin2019current} discussed the importance of e-CRM technologies while highlighting the benefits of chatbots, cloud-based solutions, and Interactive Voice Response (IVR) systems in enhancing customer service and operational efficiency. More recently, %\cite{lackovic2022predictionuserrequestcomplaint,brunello2022combinedapproachanalysisspeech} explored ASR's predictive capabilities for identifying customer requests and complaints. 
\cite{zou2021} introduced a system integrating ASR and text classification to automate transcription, track issues, and detect customer emotions in real time. 
%\cite{Parra2021} focused on developing robust ASR systems for noisy environments, emphasizing the need for denoising techniques and hybrid models to improve transcription accuracy.
\cite{kaliuta2023implementing} examines the integration of voice recognition and Natural Language Processing (NLP) in Salesforce Einstein; this work showcases AI's role in understanding customer context and sentiment to improve sales and marketing strategies. Overall, employing ASR in CRM systems enhances efficiency, accuracy, and personalization by automating the transcription of customer interactions, enriching customer profiles, and allowing sales personnel to focus on relationship building.

\smallskip

\noindent {\bf ASR Models.} Recent ASR models such as Whisper, wav2vec, wav2vec 2.0, and SpeechT5 have significantly advanced the field. Whisper \cite{radford2023robust}, developed by OpenAI, achieves near-human transcription accuracy with minimal fine-tuning using a large dataset \cite{radford2022whisper}. Wav2vec, by Facebook AI Research, uses unsupervised pre-training to achieve state-of-the-art performance \cite{schneider2019wav2vec}.
Wav2vec 2.0 improves on this with a self-supervised learning framework, showing impressive results with minimal annotated data \cite{baevski2020wav2vec20frameworkselfsupervised}. Microsoft’s SpeechT5 \cite{ao2022speecht5unifiedmodalencoderdecoderpretraining} features a unified-modal architecture for various tasks, including ASR and TTS, improving performance across these areas.

\smallskip

\noindent {\bf Weakly Supervised ASR Model Fine-tuning.} Advancements in weakly supervised learning have enhanced ASR model fine-tuning with limited labeled data. Noisy Student Training (NST) combines labeled and unlabeled data, achieving state-of-the-art results with techniques like SpecAugment \cite{Park_2020}. Almost unsupervised approaches using denoising auto-encoders also show substantial gains with minimal paired data \cite{ren2020unsupervisedtextspeechautomatic}.
Self-training methods generate and refine pseudo-labels, effectively improving ASR models in low-resource settings \cite{singh2023novelselftrainingapproachlowresource}. Data augmentation strategies, including TTS-generated synthetic data, enhance ASR performance for minority languages \cite{bartelds2023makinglittledataimproving}. Leveraging synthetic data from ASR and machine translation improves speech-to-text translation, and lightweight models are crucial for noise-resistant systems in low-resource environments \cite{Jean2022Automatic}. The LRSpeech system integrates pre-training and knowledge distillation, showing notable improvements for low-resource languages \cite{Xu2020LRSpeech}. These advancements highlight the potential of weakly supervised learning to improve ASR models with minimal labeled data.

\section{ASR-enhanced CRM Pipeline}\label{sec:pipeline}

Our desired CRM system should be suited for businesses engaged in complex customer interactions and reliant on detailed customer data for personalized service delivery. For example, the system will be valuable in the luxury goods retail sector, where personalized marketing strategies and tailored product recommendations can significantly enhance the shopping experience and customer satisfaction. This system is also suitable for sectors such as financial services,  healthcare, telecommunication, and technology retailers, etc., which offer customized solutions and continuously integrate customer feedback into product development.

To obtain insights into customer behavior, preferences, and trends, the system's priority is in obtaining accurate {\em customer portrait label} from sales personnel. Therefore, the focus of our design should be on ease of data collection, ensuring that customer information is captured accurately and in real time, without the need for manual entry. The ASR-Enhanced CRM Pipeline is illustrated in Figure~\ref{fig:framework}. After an interaction between a customer and retail personnel, the system cycles over the following steps: 

\begin{enumerate}
    \item {\bf Voice Input Capture:} Retail personnel use a mobile app equipped with a voice input device to describe customer information.

\item  {\bf Speech-to-Text Conversion:} The system processes the captured speech through ASR technology, converting it into text.

\item {\bf Data Extraction and Classification:} Natural Language Processing (NLP) techniques are used to extract key customer portrait labels from the text, which are then categorized within the CRM system.

\item {\bf Data Integration and Analytics:} The organized customer data is integrated with product data.

\item {\bf Decision Support:} Enabled by the integrated data, the CRM system offers personalized pricing, recommendations, and promotions tailored to each customer.

\item {\bf Iterative Learning and Improvement:} This continuous cycle of data collection, processing, categorization, and analytics facilitates iterative learning and improvement.
\end{enumerate}
In the rest of the paper, we shift our focus on the second step above, which is key to the CRM pipeline. 

\section{ASR Problem Definition}
Our goal is to provide an ASR model suitable for the CRM pipeline above, addressing practical challenges to provide accurate transcription into text data. The key challenge here is the lack of real domain-specific labeled data, in the form of voice-text pairs, for training an accurate ASR model. 

More formally, given a small amount of real audio data \(\mathcal{D}_r = \{(\mathbf{x}_i^r, \mathbf{y}_i^r)\}_{i=1}^{N_r}\), where \(\mathbf{x}_i^r\) is the \(i\)-th real audio sample and \(\mathbf{y}_i^r\) is its corresponding text label, with \(N_r\) being the number of real samples, and a domain-specific keyword list \(\mathcal{K} = \{k_1, k_2, \dots, k_m\}\), where \(m\) is the number of keywords, we leverage a pre-trained large language model \(M_{\text{LLM}}\) to generate synthetic text labels \(\hat{\mathbf{T}} = \{\hat{\mathbf{y}}_j^s\}_{j=1}^{N_s}\) based on \(\mathcal{K}\) and \(\{\mathbf{y}_i^r\}_{i=1}^{N_r}\), where \(N_s\) is the number of synthetic samples.

We then utilize a pre-trained Text-to-Speech (TTS) model \(M_{\text{TTS}}(\hat{\mathbf{T}})\) to generate synthetic audio \(\hat{\mathbf{X}}_s = \{\hat{\mathbf{x}}_j^s\}_{j=1}^{N_s}\) from the synthetic text \(\hat{\mathbf{T}}\). An optional step involves applying a filtering function \(\mathcal{F}\) to select high-quality synthetic data, resulting in \(\mathcal{D}_s' = \mathcal{F}(\mathcal{D}_s) = \{(\hat{\mathbf{x}}_j^s, \hat{\mathbf{y}}_j^s) \mid \text{quality}(\hat{\mathbf{x}}_j^s, \hat{\mathbf{y}}_j^s) \geq \tau\}_{j=1}^{N_s'}\), where \(\tau\) is a predefined quality threshold and \(N_s'\) is the number of filtered synthetic samples.

Finally, we fine-tune the ASR model \(M_{\text{fine}}(\mathbf{x}; \theta)\), which maps input audio \(\mathbf{x}\) to a predicted label \(\hat{\mathbf{y}}\) with parameters \(\theta\), using \(\mathcal{D}_s'\) by minimizing the loss function \(L(\mathbf{x}, \mathbf{y}; \theta) = \text{CrossEntropy}(M_{\text{fine}}(\mathbf{x}; \theta), \mathbf{y})\). The fine-tuned optimal ASR model $M$ will be deployed in industrial applications.

\section{Methodology}

\subsection{Weak Supervision ASR}\label{sec:method}

\begin{figure}[!htbp] % 图片环境
    \centering % 图片居中
    \includegraphics[width=\textwidth]{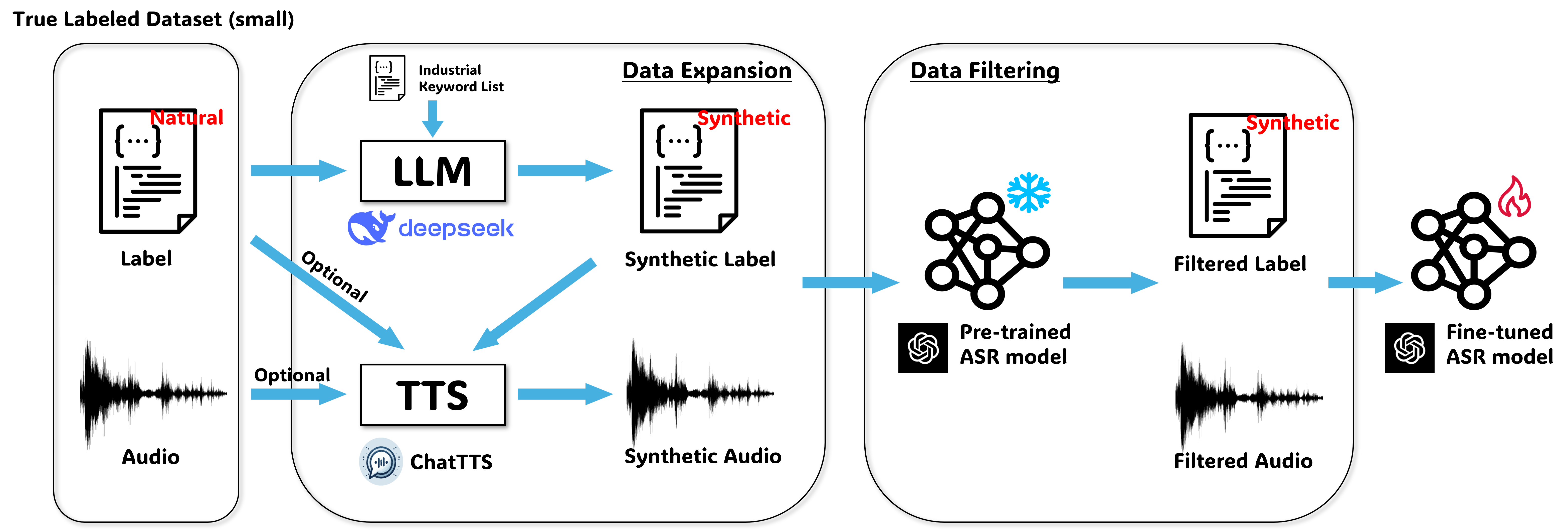} %
    \caption{The detailed main framework of the ASR fine-tuning method, which combines the LLM (DeepSeek) and TTS model (ChatTTS), includes two main parts: data expansion and data screening.} % 图片标题
    \label{fig:framework} % 图片标签，方便在文中引用
\end{figure}

Figure~\ref{fig:framework} details the specific process by which we build a dedicated ASR model for a specific retail company and apply it to the corresponding CRM system. It consists of two main parts: {\em data expansion} and {\em data filtering}.

We use the labeled speech data of real business scenarios the industry provides as the input framework, aiming to generate a large amount of high-quality labeled synthetic data. First, we use LLM (DeepSeek V2 \cite{deepseekv2}) to generate labels for all synthetic data. The way to generate labels is to imitate the expression in the original data labels. In this process, we provide the language model with a keyword list containing most of the industry's professional terms. This keyword list is mainly obtained by crawling relevant industry data on social media and manually cleaning and filtering. After generating labels for all synthetic data, we use an advanced TTS model (ChatTTS\footnote{https://github.com/2noise/ChatTTS}) to complete the speech synthesis task, taking the relevant source data and the interference components in the original data (such as dialects, accents, etc.) as one of the input parameters, simulating various complex situations encountered in real scenarios, and thus obtaining a large amount of synthetic speech data and its corresponding real labels.

After expanding the data to obtain a large dataset, we need to further conduct an additional data filtering process to ensure the high quality of the synthetic data. First, we use a pre-trained ASR model without fine-tuning (here we use whisper-large-v2) to infer the corresponding tags for all synthetic data. We then compare these inferred tags with the originally generated tags to calculate the CER metric. We set a threshold for this metric and exclude all data that do not meet this criterion. The remaining synthetic speech data and tags will form the final synthetic dataset for fine-tuning the ASR model. Also, in the fine-tuning stage of the ASR model, we consider using LoRA fine-tuning instead of SFT fine-tuning \cite{han2024parameter} to reduce computing resource overhead while minimizing performance loss.

\subsection{Integrated Error Rate}\label{sec:IER}

Traditional speech recognition model indicators include word error rate (WER) \cite{park2024automatic} and character error rate (CER) \cite{wigington2017data}, which are respectively applicable to monolingual speech recognition tasks in different languages. For example, English recognition is more suitable for WER, while Chinese recognition is more suitable for CER. However, in some industrial scenarios, such as the sales process of luxury brands, the communication between customers and sales staff usually retains the original pronunciation of the brand without translating it into a common language. Therefore, the above two indicators cannot be used as a good evaluation criterion for such multilingual mixed speech recognition tasks.

To address this problem, we propose the Integrated Error Rate (IER). IER combines the advantages of WER and CER and evaluates the accuracy of Chinese phrases and single Chinese characters in detail. In addition, we introduce the concept of keyword recognition and integrate it into IER to evaluate the accuracy of ASR models in recognizing unknown keywords.

The label of speech data $X$ can be considered to consist of three parts: words $\mathcal{W}$, characters $\mathcal{C}$ and keywords $\mathcal{S}$, that is, $X = \bigcup_{i=1}^{n} \{W_i, C_i, S_i\}$. To understand the composition of the sentence in detail and calculate the score comprehensively, we used the Chinese word segmentation tools. Since the potential disagreement in the semantics of Chinese word segmentation may lead to deviations in the word segmentation results, we combined the current mainstream Chinese word segmentation tools jieba\footnote{https://pypi.org/project/jieba/} and HanLP\footnote{https://github.com/hankcs/HanLP} \cite{he-choi-2021-stem} to divide the text into the above three parts and calculate the IER results.

We define the integrated error rate (IER) as follows:

\[
\scriptsize
\begin{aligned}
\text{IER} = \frac{1}{M} \sum_{m=1}^{M} \Bigg[ & \left( \frac{|\mathcal{W}_m|}{|\mathcal{W}_m| + |\mathcal{C}_m| + |\mathcal{S}_m|} \right) \cdot \left( \frac{1}{|\mathcal{W}_m|} \sum_{i=1}^{|\mathcal{W}_m|} \delta_i \right) \\
& + \left( \frac{|\mathcal{C}_m|}{|\mathcal{W}_m| + |\mathcal{C}_m| + |\mathcal{S}_m|} \right) \cdot \left( \frac{1}{|\mathcal{C}_m|} \sum_{j=1}^{|\mathcal{C}_m|} \epsilon_j \right) \\
& + \left( \frac{|\mathcal{S}_m|}{|\mathcal{W}_m| + |\mathcal{C}_m| + |\mathcal{S}_m|} \right) \cdot \left( \frac{1}{|\mathcal{S}_m|} \sum_{k=1}^{|\mathcal{S}_m|} \zeta_k \right) \Bigg]
\end{aligned}
\]

Here, \(M\) is the total number of segmentation methods, \(m\) is the current segmentation method, and \(|\mathcal{W}_m|\), \(|\mathcal{C}_m|\), and \(|\mathcal{S}_m|\) represent the number of words, characters, and special terms in the current segmentation method, respectively. The indicator functions \(\delta_i\), \(\epsilon_j\), and \(\zeta_k\) represent the errors for corresponding words, characters, and special terms. Specifically, \(\delta_i = 1\) if the \(i\)-th word is incorrect, otherwise \(\delta_i = 0\); \(\epsilon_j = 1\) if the \(j\)-th character is incorrect, otherwise \(\epsilon_j = 0\); and \(\zeta_k = 1\) if the \(k\)-th keyword is incorrect, otherwise \(\zeta_k = 0\).

Due to the potential overlap between WER and CER, we implemented a de-duplication process during calculation. Specifically, for each incorrect word, we count it only once in the word error rate and do not double-count it in the character error rate. The following is an analysis of a real case:

\begin{figure}[!htbp] % 图片环境
    \centering % 图片居中
    \includegraphics[width=\textwidth]{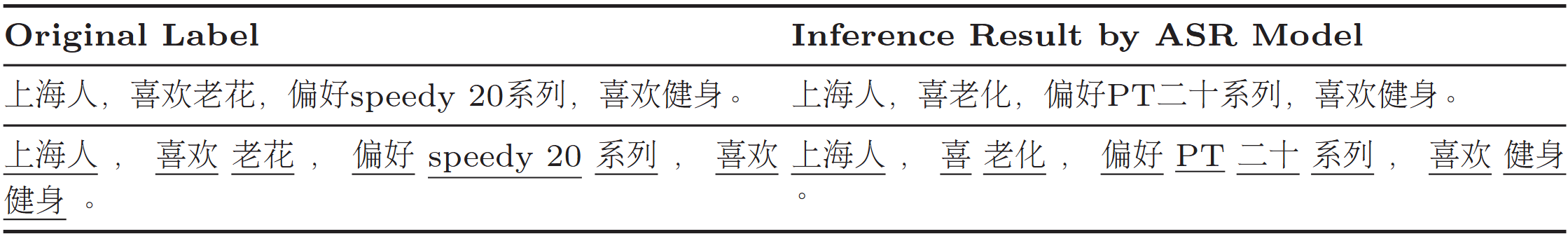} %
    \caption{Original true label content of a fictitious label and the version inferred by the ASR model. The second row of data is the result through HanLP word segmentation.} % 图片标题
    \label{fig:IERExample} % 图片标签，方便在文中引用
\end{figure}

The inference result in Figure~\ref{fig:IERExample} shows 1 character error: ``xi", 1 word error: ``laohua", and 1 keyword error: ``PT" and ``ershi". ``Speedy 20" is an inseparable part in the keyword lists but was split into two incorrect words in the inference results. However, by considering the positions of the correct words before and after, we summarize this as containing only one error. Since this method requires a list of industry-specific keywords and involves complex differences in word segmentation strategies, we manually verified all automatically calculated indicators to ensure their theoretical correctness. This approach allows us to more accurately evaluate the overall performance of ASR models in multilingual environments and specific industries.

\section{Experimental Settings}\label{sec:exp}

\subsection{Datasets}

Three original datasets are mentioned in this paper, all provided by Atom Intelligence Group, namely \textbf{GUCCI100}, \textbf{LV100} and \textbf{Test Set}. Our well-trained sales staff record these datasets in real luxury business scenarios. They are entered according to some fixed language expressions and manually annotated to ensure the details and accuracy of the labels. Each voice tag may contain a rough user portrait obtained through observations of sales staff communicating with customers, including information such as age and preferences. The following is an approximate example:

\begin{figure}[!htbp] % 图片环境
    \centering % 图片居中
    \includegraphics[width=\textwidth]{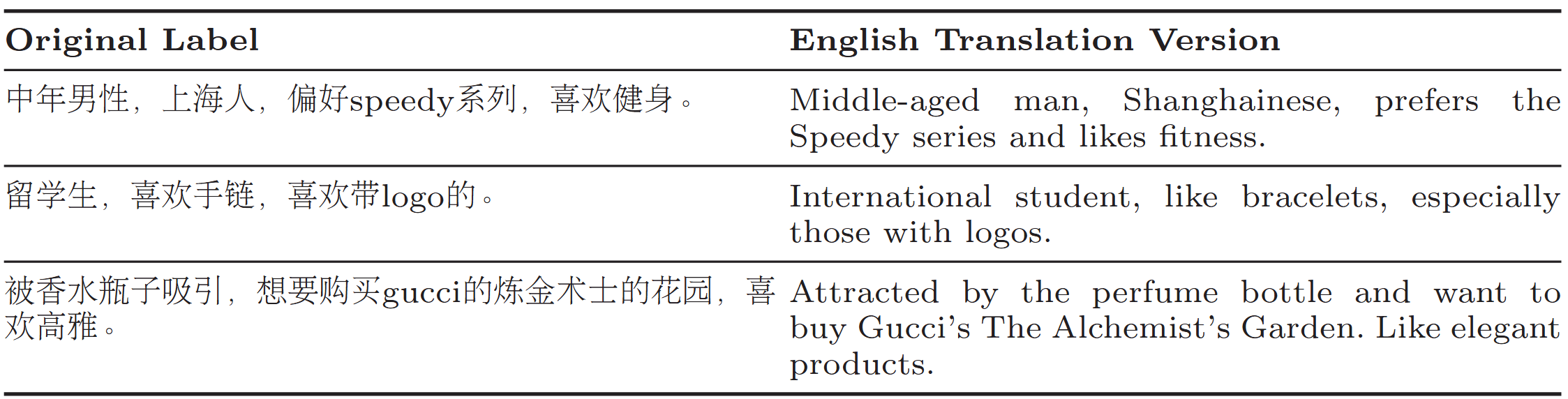} %
    \caption{The true labels and English translations. These examples are fictitious, they just mimic the label format and are not real data from actual scenarios.} % 图片标题
    \label{fig:example} % 图片标签，方便在文中引用
\end{figure}

\begin{table}[b]
\centering
\scriptsize
\begin{tabular}{llll}
\hline
\textbf{Dataset}  & \textbf{Samples} & \textbf{Duration (seconds, avg $\pm$ std)} & \textbf{Audio length (minutes)} \\ \hline
GUCCI100     & 100               & 8.58$\pm$4.07         & 14.87                     \\
LV100        & 100               & 8.92$\pm$3.39         & 14.31                     \\
Test Set     & 1000              & 8.74$\pm$3.34         & 145.46                    \\
LVChatTTS    & 10000             & 8.81$\pm$3.21         & 1468.98                   \\
GUCCIChatTTS & 10000             & 8.98$\pm$3.12         & 1496.14                   \\ \hline
\end{tabular}
\caption{Data Statistics}
\label{tab:dat_summ}
\end{table}

\begin{table}[t]
\centering
\scriptsize
\begin{tabular}{lll}
\hline
\textbf{Category} & \textbf{Samples} & \textbf{Example}                 \\ \hline
SERIES   & 408     & objets nomades          \\
TYPE     & 273     & decorations       \\
BRAND    & 92      & balenciaga              \\
MATERIAL & 42      & empreinte               \\
NICKNAME & 42      & bucket bag        \\
LINES    & 19      & monogram                \\
SOCIAL   & 3       & value preservation \\ \hline
\end{tabular}
\caption{Keywords Statistics}
\label{tab:kw_summ}
\end{table}

GUCCI100 and LV100 each contain 100 real data information from two luxury brands, while Test Set is a labeled dataset combining different luxury brands. In the experimental part of this paper, the Test Set is mainly used for testing the fine-tuned ASR models and does not participate in the construction of any fine-tuning datasets. Detailed statistics for these datasets are presented in Table~\ref{tab:dat_summ}. In addition to these datasets, the keywords list \(\mathcal{K}\), detailed in Table~\ref{tab:kw_summ}, features seven categories used to generate synthetic training samples. Note that the individual audio recordings for each keyword is absent. Relying on these two small datasets and a keywords list $\mathcal{K}$, we used the paradigm framework mentioned in the paper to generate 10,000 synthetic datasets for each brand. These two datasets will be used as independent datasets in the model fine-tuning process and will be merged to form a final version containing 20,000 data points for the model fine-tuning process.

\subsection{Evaluation Metrics}

We mainly used CER and WER as indicators in the experiment. In addition, since this task involves multi-language translation (classified as Chinese and other languages), we calculate the recognition scores of the two languages separately, including CER\_cn, CER\_oth, WER\_cn, and WER\_oth. At the same time, we use the integrated error rate (IER) mentioned above as an additional evaluation indicator to evaluate the model performance independently.

\subsection{Pseudo Label Generation}

We use a small labeled dataset and a keywords list $\mathcal{K}$ to generate pseudo-labels for training. The objective is to ensure diversity in the synthetic dataset by covering as many keywords in $\mathcal{K}$, and as many styles of the refernce sentences as possible, thereby producing varied audio samples. To achieve this, we randomly sample \textit{s} sentences from real data and \textit{i} keywords from each category in $\mathcal{K}$. If a category contains fewer than \textit{i} keywords, we select all available keywords in that category. These sentences and keywords are then used as input prompts for DeepSeek V2 to generate the pseudo-labels. In this study, \textit{s} is set to 5 and \textit{i} is set to 8. 

We use the following prompt to generate pseudo-labels, where the content inside curly braces represents variables. The prompt includes SENTENCE, which is a concatenation of real labels separated by a newline character, and other variables separated by a comma pattern:

\begin{mdframed}
\scriptsize
\noindent \textbf{System prompt}:
\begin{lstlisting}
Please refer to the following examples and consider the existing product information. Take a deep breath and think carefully, can you provide additional examples? These examples should closely align with the original samples. For product-related information, please use the product list we have provided.
\end{lstlisting}

\noindent \textbf{User prompt}:
\begin{lstlisting}
Below is the product information for your reference:
Attribute: {SOCIAL}; Brand: {BRAND}; Pattern: {LINES};
Material: {MATERIAL}; Product: {NICKNAME}; Series: {SERIES}; 
Type: {TYPE};
Here are some examples: {SENTENCE}
Based on the template in the examples, please generate {s} new sentences. The content and style should be aligned with the examples.
\end{lstlisting}
\end{mdframed}

\subsection{Experimental Environment/Startup}
We use the power of LLMs by calling the API of DeepSeek V2. The speech data was generated using the current mainstream open-source TTS model ChatTTS. The fine-tuning of the ASR model refers to the open-source LoRA fine-tuning solution on GitHub, which can be viewed here. The important parameters include the number of fine-tuning epochs is 5, the batch size is 4, the learning rate is 1e-3, and the maximum audio length is 30 seconds (all label data does not exceed this limit). We use an NVIDIA RTX A6000 GPU with 40G VRAM.

We will use each of the above fine-tuning dedicated datasets to fine-tune three different versions of the Whisper model and perform result testing and metric calculation on the test set. We refer to the GitHub open-source solution\footnote{https://github.com/yeyupiaoling/Whisper-Finetune} for LoRA fine-tuning methods and codes. Due to the access restrictions of HanLP and the uncertainty of the word segmentation strategy mentioned in the method, we invested a lot of human resources to verify the calculation results of IER. In the experiment, we only calculated the IER indicator for the best-performing models after fine-tuning among the three versions. We reduced the amount of data in the test set to 200.

\section{Experimental Results}

Table \ref{tab:result1} shows the inference results of three different versions of the \textbf{Whisper} model (whisper-medium, whisper-large-v2, whisper-large-v3) after fine-tuning on different datasets mentioned in this paper, where some of the best results are shown in bold.

\begin{table}[!htbp]
    \centering
    \scriptsize %
    \resizebox{\textwidth}{!}{ % 调整表格宽度以适应页面
    \begin{tabularx}{\textwidth}{lXXXXXX}
        \toprule
        \textbf{Model} & \textbf{CER} & \textbf{CER\_cn} & \textbf{CER\_oth} & \textbf{WER} & \textbf{WER\_cn} & \textbf{WER\_oth} \\ 
        \midrule
        whisper-medium & 0.54125 & 0.50068 & 0.70524 & 1.48446 & 0.98699 & 0.94388 \\ 
        medium-GUCCI100 & 0.38658 & 0.30772 & 0.45721 & 0.85850 & 0.69 & 0.36364 \\ 
        medium-GUCCIChatTTS & 0.21169 & 0.11968 & 0.38572 & 1.30250 & 0.72773 & 0.73488 \\
        medium-LV100 & 0.42269 & 0.34472 & 0.67785 & 0.69223 & 0.47222 & 0.58127 \\ 
        medium-LVChatTTS & 0.21234 & 0.14897 & 0.34662 & 1.13762 & 0.70 & 0.88423 \\ 
        medium-GUCCCI\&LV & 0.20007 & 0.19243 & 0.22237 & 0.77381 & 0.68225 & 0.73338 \\ 
        \midrule
        whisper-large-v2 & 0.11958 & 0.06540 & 0.24480 & 1.56604 & 0.58 & 0.36471 \\ 
        v2-GUCCI100 & 0.12333 & 0.06677 & 0.18500 & 0.70142 & 0.35 & 0.35498 \\ 
        v2-GUCCIChatTTS & 0.08795 & 0.07739 & 0.14031 & 0.71664 & \textbf{0.33921} & 0.44834 \\ 
        v2-LV100 & 0.07996 & 0.05031 & \textbf{0.12779} & 0.99662 & 0.44989 & 0.57324 \\ 
        v2-LVChatTTS & 0.07743 & 0.05552 & 0.17980 & 0.83076 & 0.42877 & \textbf{0.29445} \\ 
        v2-GUCCCI\&LV & \textbf{0.07390} & \textbf{0.04593} & 0.13383 & \textbf{0.62264} & 0.44 & 0.31461 \\ 
        \midrule
        whisper-large-v3 & 0.18793 & 0.09832 & 0.22471 & 1.17422 & 0.82 & 0.99314 \\ 
        v3-GUCCI100 & 0.14223 & 0.07774 & 0.16648 & 1.22439 & 0.897 & 0.88442 \\ 
        v3-GUCCIChatTTS & 0.08732 & 0.07002 & 0.09956 & 1.04436 & 0.66793 & 0.98092 \\ 
        v3-LV100 & 0.11339 & 0.11042 & 0.13398 & 1.11147 & 0.87 & 0.93732 \\ 
        v3-LVChatTTS & 0.11271 & 0.10887 & 0.16643 & 0.90742 & 0.72 & 0.80452 \\ 
        v3-GUCCCI\&LV & 0.09332 & 0.07741 & 0.13346 & 0.80201 & 0.69 & 0.78147 \\ 
        \bottomrule
    \end{tabularx}}
    \caption{The performance of the three base models after fine-tuning on different training sets. Model names without any special suffixes are the native model capability results. GUCCI100/LV100 are the results of fine-tuning the models using the real labeled datasets of the two brands, respectively. GUCCIChatTTS/LVChatTTS are the results of using the high-quality datasets generated by our proposed framework. GUCCCI\&LV are the results of merging two synthetic datasets.} % 添加表格标题
    \label{tab:result1} % 添加表格标签
\end{table}

Since the main language used in our fine-tuning and test datasets is Chinese, in theory, CER-related indicators are more suitable for evaluating Chinese speech recognition tasks. From this, we can see that the whisper-large-v2 basic model achieved the best CER performance after fine-tuning the final dataset after merging the two synthetic datasets of GUCCI and LV, and the latest whisper-large-v3 model did not have the best performance as expected. In addition, for indicators of other languages (CER\_oth and WER\_oth), whisper-large-v2 did not achieve the best performance indicators. This may be because the false-labeled speech generated by ChatTTS deviates from the pronunciation of other languages in real scenarios, resulting in a large bias in the synthetic dataset itself.

Similarly, we calculated the IER of the best fine-tuned versions of the three models and manually verified the indicators. The results are shown in Table \ref{tab:result2}. As expected, the best fine-tuned version of whisper-large-v2 achieved the best results on this indicator. However, the performance of whisper-large-v3 was unexpected again, and it did not even exceed the model fine-tuned by whisper-medium. We will analyze the reasons for this kind of accident in detail later.

\begin{table}[!htbp]                                
    \centering
    \scriptsize %
    \begin{tabularx}{0.5\textwidth}{l>{\centering\arraybackslash}X}
        \toprule
        \textbf{Model} & \textbf{IER} \\ 
        \midrule
        whisper-medium-GUCCCI\&LV & 0.3640  \\ 
        \midrule
        whisper-large-v2-GUCCCI\&LV & 0.2227  \\ 
        \midrule
        whisper-large-v3-GUCCIChatTTS &  0.3962  \\ 
        \bottomrule
    \end{tabularx}
    \caption{The values of the IER indicator for the models with the best fine-tuning results for the three versions (based on CER).} % 添加表格标题
    \label{tab:result2} % 添加表格标签
\end{table}
\section{Detailed Analysis}

\subsection{Case Study}

\subsubsection*{Why whisper-large-v3 performs worse?}
\begin{figure}[!htbp] % 图片环境
    \centering % 图片居中
    \includegraphics[width=\textwidth]{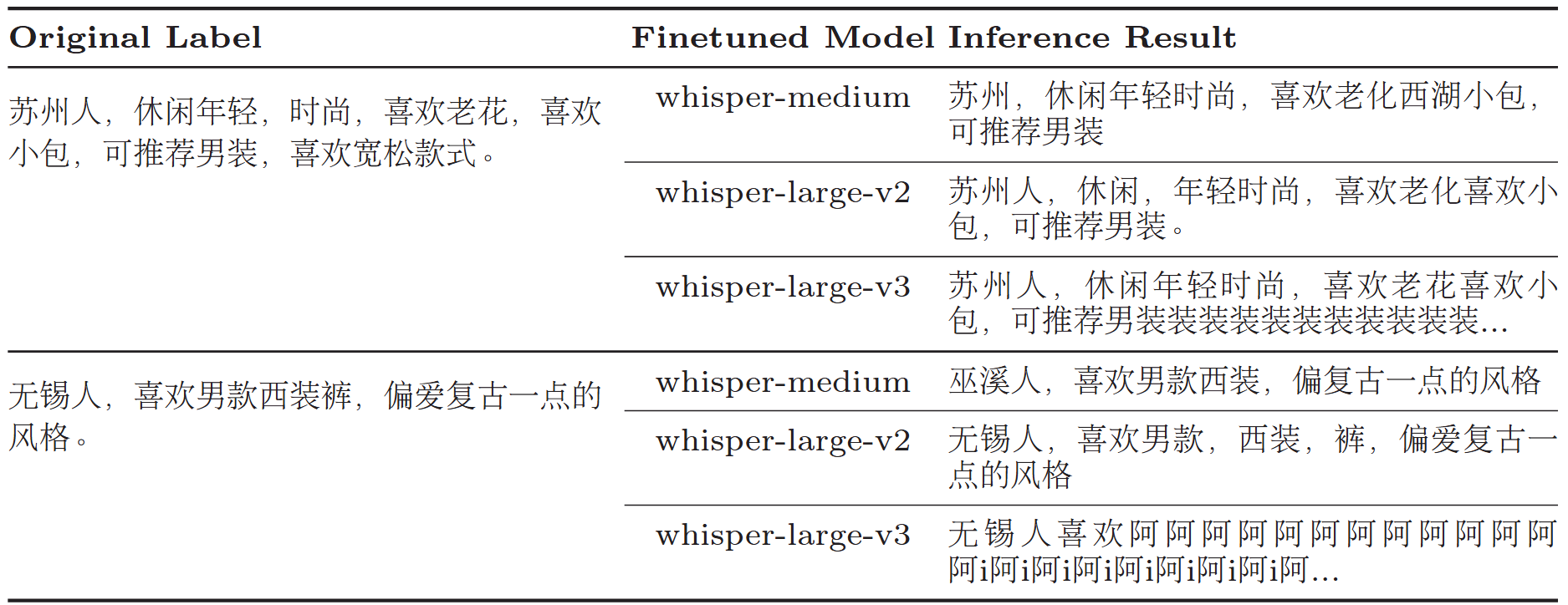} %
    \caption{The original labels and inference results by different models in the test set. (Data anonymization has been completed)} % 图片标题
    \label{fig:whisper-large-v2} % 图片标签，方便在文中引用
\end{figure}

We checked the inference results of the whisper-large-v3 fine-tuned version of the model. As shown in Table \ref{fig:whisper-large-v2}, in the second half of the inference task, the results of whisper-large-v3 showed abnormal repetition of single Chinese characters/phrases, resulting in poor results in various indicators. However, whisper-large-v2 has achieved almost zero error translation performance after fine-tuning.

The performance of the whisper-large-v3 baseline model after fine-tuning is not as stable as that of whisper-large-v2 in most cases. This, in fact, has been a well-recognised problem with whisper-large-v3 (see OpenAI official forum). We suspect that the increase in the number of parameters of the model may cause bias in the synthetic data to be learned as data features by the model after fine-tuning, which affects the model's performance.

\subsection{Selection of Num\_epoch and CER Threshold}

\begin{figure}[!htbp]
    \centering
    \begin{subfigure}[b]{0.48\textwidth}
        \centering
        \includegraphics[width=\textwidth, height=4cm]{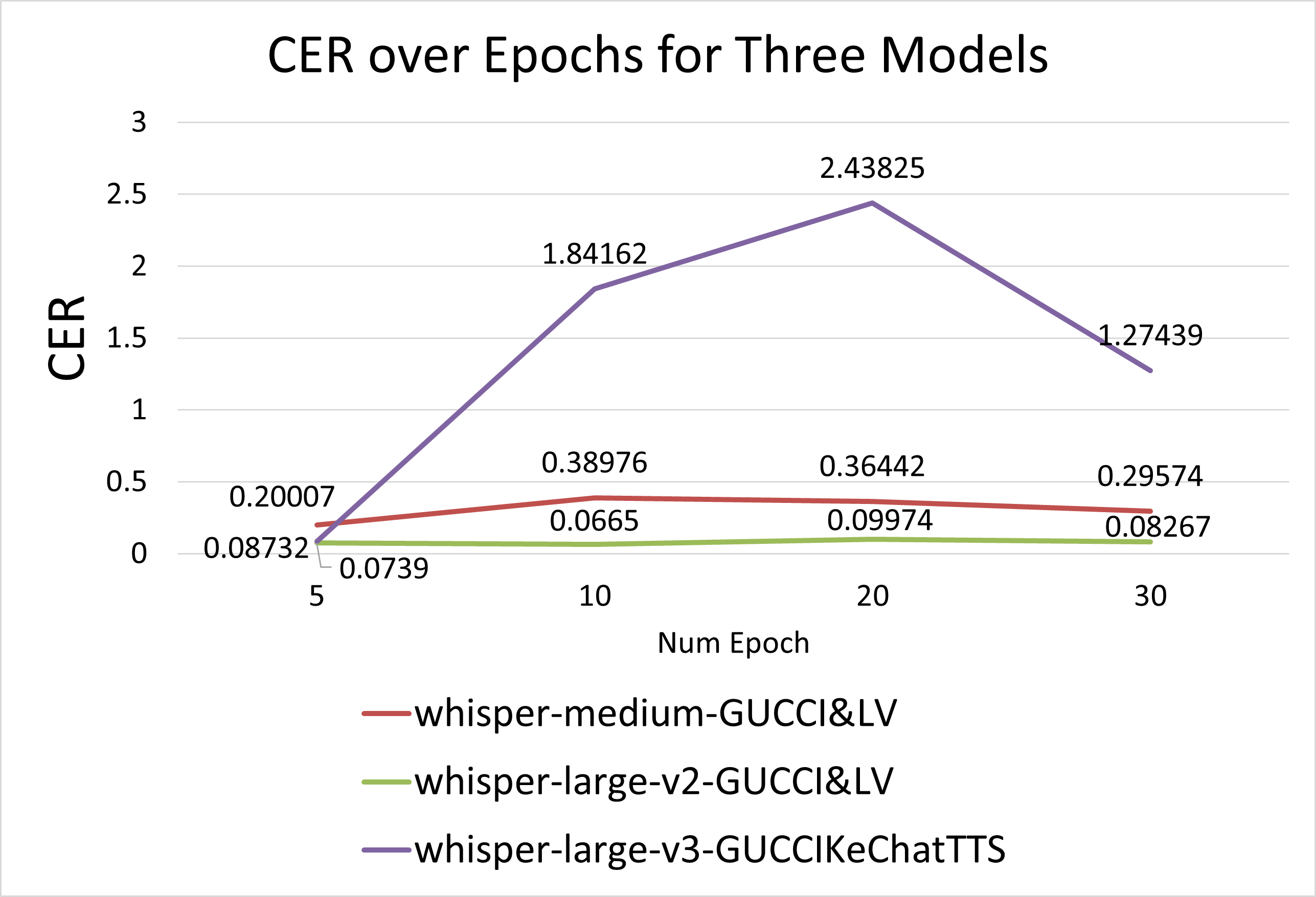} % 替换为第一张图片的路径
        \caption{CER indicator results after further fine-tuning the three models with 10, 15, 20, 30 epoch.}
        \label{fig:first_image}
    \end{subfigure}
    \hfill
    \begin{subfigure}[b]{0.48\textwidth}
        \centering
        \includegraphics[width=\textwidth, height=4cm]{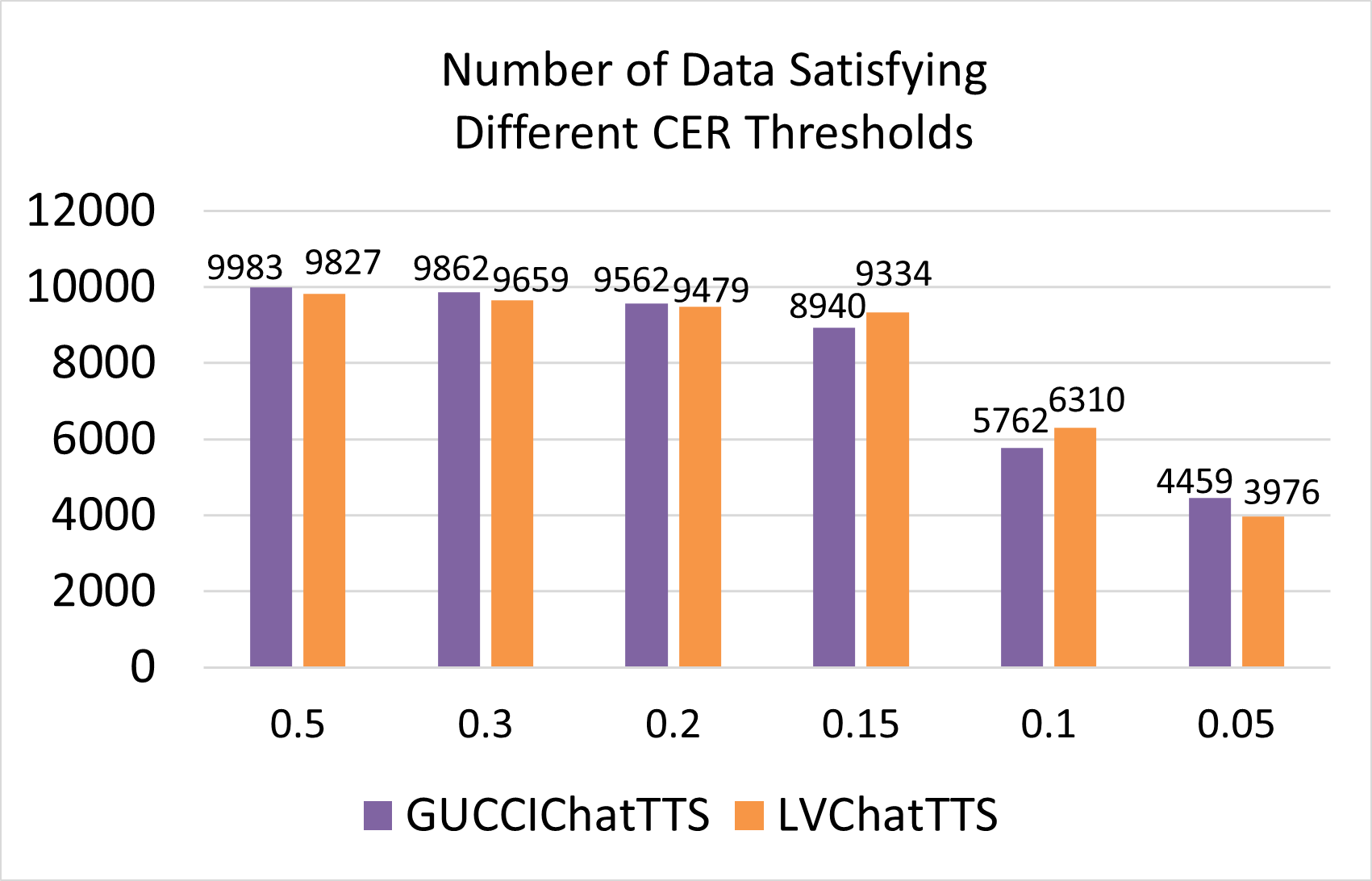} % 替换为第二张图片的路径
        \caption{Number of remaining data in the two synthetic datasets after filtering with different CER indicators.}
        \label{fig:second_image}
    \end{subfigure}
    \caption{Results of additional experiments.}
    \label{fig:combined_images}
\end{figure}

Our experiments determined that each model's optimal number of fine-tuning epochs is 5. To further validate this conclusion, we conducted additional experiments. Sub-figure \ref{fig:first_image} of Figure \ref{fig:combined_images} shows the results of further fine-tuning the three best-performing models while keeping other parameters constant and calculating their CER metrics. As the number of epochs increased, whisper-large-v3 was significantly affected, with its performance deteriorating substantially. The other two models showed fluctuations but generally trended downward, possibly due to overfitting. Therefore, we chose 5 epochs as one of the parameters for fine-tuning the ASR model.

In addition, a CER threshold is determined to screen the dataset. In sub-figure \ref{fig:second_image}, we show the remaining data in the two synthetic datasets under different CER boundaries. By weighing the amount and quality of data, we choose 0.15 as the filtering boundary. By manually randomly checking the synthetic data, we found that their quality is within an acceptable range.

\section{Conclusion and Limitation}

We proposed a weak supervision ASR model fine-tuning framework suitable for the industry-specific CRM system, including detailed data expansion and filtering processes. The ASR model fine-tuned by this framework is competent for its task and performs well. 

While this paradigm framework can generate results from a small amount of industry-specific labeled data, it is sometimes limited by the training data, causing it to overlook parts of the original prompt or include illegal text in the output (depending on the text review capabilities of the large model platform). In rare cases, due to language models' autoregressive and generative nature, the ``fake labels'' generated by the LLM may be semantically inconsistent with the original prompt, and the TTS model may produce partially unusable audio. These issues will be addressed in our future work. Additionally, using multiple models during data construction and fine-tuning can introduce some bias and error accumulation.

\medskip

\noindent {\bf Acknowledgements.} This project is supported by Atom Intelligence Group, which provides the necessary data information and computing resources. The method described in the paper has been tested in Atom Intelligence Group.

\bibliographystyle{splncs04}
\bibliography{mybib}
%

% \end{CJK}
\end{document}